\documentclass[12pt]{article}
\usepackage{epsfig}

\newcommand{\mysection}{\setcounter{equation}{0}\section}

\def\beq{\begin{equation}}
\def\eeq{\end{equation}}
\def\beqa{\begin{eqnarray}}
\def\eeqa{\end{eqnarray}}

\newlength{\dinwidth} \newlength{\dinmargin}
\begin{document}

\begin{center}
{\Large \bf Higher-order QCD corrections  for the
$W$-boson transverse momentum distribution}
\end{center}
\vspace{2mm}
\begin{center}
{\large Nikolaos Kidonakis$^a$ and Richard J. Gonsalves$^b$}\\
\vspace{2mm}
${}^a${\it Kennesaw State University,  Physics \#1202,\\
1000 Chastain Rd., Kennesaw, GA 30144-5591, USA} \\
\vspace{2mm}
${}^b${\it Department of Physics, University at Buffalo,
The State University of New York, \\ Buffalo, NY 14260-1500, USA}
\end{center}

\begin{abstract}
We present results for $W$-boson production at large transverse momentum
at LHC and Tevatron energies. We calculate complete next-to-leading-order (NLO)
QCD corrections and higher-order soft-gluon corrections to the differential cross section.
The soft-gluon contributions are resummed at next-to-next-to-leading-logarithm
(NNLL) accuracy via the two-loop soft anomalous dimensions.
Both NLO and approximate next-to-next-to-leading-order (NNLO) $p_T$ distributions are
presented. Our numerical results are in good agreement with recent data from the LHC.
\end{abstract}

\mysection{Introduction}

The production of $W$ bosons with large transverse momentum, $p_T$, has been
observed and analyzed at the Tevatron over the past two decades,
and significantly higher event rates
have been observed as expected at the LHC over the past couple of years.
This Standard Model process is a
background to Higgs production and new physics and thus it is important
to have accurate theoretical predictions to exploit fully the large number
of events at the LHC. The $p_T$ distribution falls rapidly with increasing
$p_T$, spanning several orders of magnitude over accessible regions at
hadron colliders.
High-$p_T$ $W$ production has a clean experimental signature when the $W$
decays to leptons, and solid predictions are needed to
reduce uncertainties in precision measurements of the $W$ mass and decay width.
The charged leptons in complementary processes involving $Z$ bosons
can be measured with somewhat higher resolution than the neutrino,
but the observed event rate for $W$ bosons at the LHC is as much as a factor
of ten larger than that for $Z$ bosons.
Precise calculations for $W$ production at large $p_T$ are also
needed to identify signals of possible new physics, such as new
gauge bosons, which may enhance the $p_T$ distribution at large $p_T$.

At leading order (LO) in the strong coupling
$\alpha_s$, a $W$ boson can be produced with large $p_T$ by recoiling
against a single parton which decays into a jet of hadrons.
The LO partonic processes for $W$ production at large $p_T$ are
$q g \rightarrow W q$ and $q {\bar q} \rightarrow W g$.

The NLO corrections arise from one-loop parton
processes with a virtual gluon, and real radiative processes with
two partons in the final state.
The NLO corrections to the cross section for $W$ production at large
$p_T$ were calculated in \cite{AR,gpw} where complete analytic expressions
were provided.
Numerical NLO results for production at the  Tevatron were also presented in
Refs. \cite{AR,gpw}.
The predictions are consistent with the
data from the CDF \cite{CDF} and D0 \cite{D0} collaborations.
The NLO corrections enhance the
differential distributions in $p_T$ of the $W$ boson
and they reduce the factorization and renormalization
scale dependence.

Beyond NLO, it is possible to calculate contributions from
the emission of  soft gluons.
These corrections can be formally resummed and they were first calculated to
next-to-leading-logarithm (NLL) accuracy in \cite{NKVD}.
Approximate NNLO corrections derived from the resummation were used in \cite{NKASV}
and were shown to provide enhancements and a further reduction
of the scale dependence. Numerical results were presented
for the Tevatron in \cite{NKASV} and for the LHC at 14 TeV energy in
Ref. \cite{GKS}. In this paper we extend the resummation to
next-to-next-to-leading-logarithm (NNLL) accuracy (see also \cite{NKRGdpf}).
A related study using soft-collinear effective theory (SCET) has
recently appeared in \cite{BLS}.

In the next section we briefly review the NLO calculation and present numerical
results for the $p_T$ distribution of the $W$ at the LHC and the Tevatron.
Section 3 discusses NNLL resummation for the soft-gluon corrections.
In section 4 we derive approximate NNLO expressions from the NNLL resummation
and we present approximate NNLO $p_T$ distributions for the $W$ boson at the
LHC and the Tevatron. We conclude in Section 5.

\mysection{NLO results}

We start with the leading-order contributions to $W$ production at
large $p_T$ with a single hard parton in the final state.
The two contributing sub-processes are
$$q(p_a) + g(p_b) \longrightarrow W(Q) + q(p_c)$$
and
$$q(p_a) + {\bar q}(p_b) \longrightarrow W(Q) + g(p_c) .$$
We define the kinematic variables $s=(p_a+p_b)^2$, $t=(p_a-Q)^2$,
$u=(p_b-Q)^2$, and $s_4=s+t+u-Q^2$.
At the partonic threshold, where there is no available energy for
additional radiation, $s_4 \rightarrow 0$.
The partonic cross sections are singular in this limit, but the divergences
are integrable when averaged over the parton distributions in the colliding
hadrons.

\begin{figure}
\begin{center}
\includegraphics[width=10cm]{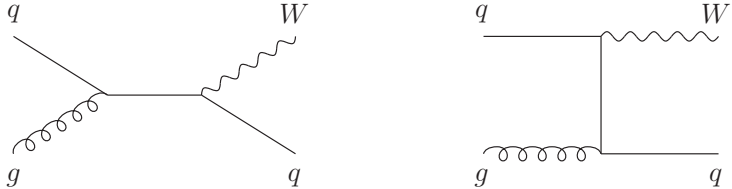}
\caption{LO diagrams for the process $qg \rightarrow Wq$.}
\label{qgqWdiag}
\end{center}
\end{figure}

\begin{figure}
\begin{center}
\includegraphics[width=10cm]{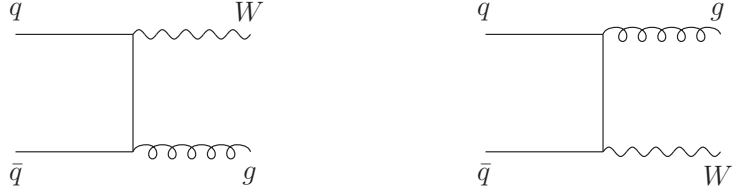}
\caption{LO diagrams for the process $q{\bar q} \rightarrow Wg$.}
\label{qqbWgdiag}
\end{center}
\end{figure}

The LO diagrams are shown in Figs. \ref{qgqWdiag} and \ref{qqbWgdiag}.
The LO differential cross section for the $qg \rightarrow Wq$ process is
\beq
E_Q \frac{d\sigma^B_{qg\rightarrow Wq}}{d^3Q}
=F^B_{qg \rightarrow Wq} \, \delta(s_4) \, ,
\eeq
where
\beqa
F^B_{qg \rightarrow Wq} &=& \frac{\alpha \,
\alpha_s(\mu_R^2)C_F}{s(N_c^2-1)}
A^{qg} \, \sum_{f} |L_{ff_a}|^2 \, ,\\
A^{qg} &=& - \left(\frac{s}{t}+\frac{t}{s}+\frac{2uQ^2}{st}\right) \, ,
\nonumber
\eeqa
with $\mu_R$ the renormalization scale, $L$ the left-handed couplings of the
$W$ boson to the quark line, and $f$ the quark flavor.
Also $C_F=(N_c^2-1)/(2N_c)$ with $N_c=3$ the number
of colors.

For the process $q{\bar q} \rightarrow Wg$ the LO result is
\beq
E_Q \frac{d\sigma^B_{q {\bar q}\rightarrow Wg}}{d^3Q}
=F^B_{q{\bar q} \rightarrow Wg} \, \delta(s_4) \, ,
\eeq
where
\beqa
F^B_{q{\bar q} \rightarrow Wg} &=&\frac{\alpha \, \alpha_s(\mu_R^2)C_F}{sN_c}
A^{q\bar q}\,  |L_{f_bf_a}|^2 \, , \\
A^{q\bar q} &=& \frac{u}{t}+\frac{t}{u}+\frac{2Q^2s}{tu} \, .
\nonumber
\eeqa

The complete NLO corrections were derived in \cite{AR,gpw}.
The virtual corrections involve
ultraviolet divergences which renormalize $\alpha_s$ and make it
depend on the renormalization energy scale which we set to be ${\sim}p_T$.
Both the real and the virtual corrections display soft and collinear
divergences which arise from the masslessness of the gluons and the
zero-mass approximation for the quarks.
The soft divergences cancel between real and virtual processes while
the collinear singularities are factorized in a process-independent
manner and absorbed into factorization-scale-dependent parton
distribution functions.

The complete NLO corrections to the LO differential cross section can be
written as a sum of two terms
\beq
E_Q\,\frac{d\hat{\sigma}^{(1)}_{f_af_b{\rightarrow}W(Q)+X}}{d^3Q}=
\alpha_s^2(\mu_R^2) \left[\delta(s_4) \, B(s,t,u,\mu_R)
+C(s,t,u,s_4,\mu_F) \right]
\eeq
with $\mu_F$ the factorization scale.
The coefficient functions $B$ and $C$ depend on the parton flavors.
$B(s,t,u,\mu_R)$ is the sum of virtual corrections and of singular terms
${\sim}\delta(s_4)$ in the real radiative corrections.
$C(s,t,u,s_4,\mu_F)$ is from real emission processes away from $s_4=0$.
The NLO corrections are crucial in reducing theoretical uncertainties and
thus making more meaningful comparisons with experimental data for
$W$ production at the Tevatron \cite{CDF,D0} and the LHC \cite{ATLAS}
at large transverse momentum.

\begin{figure}
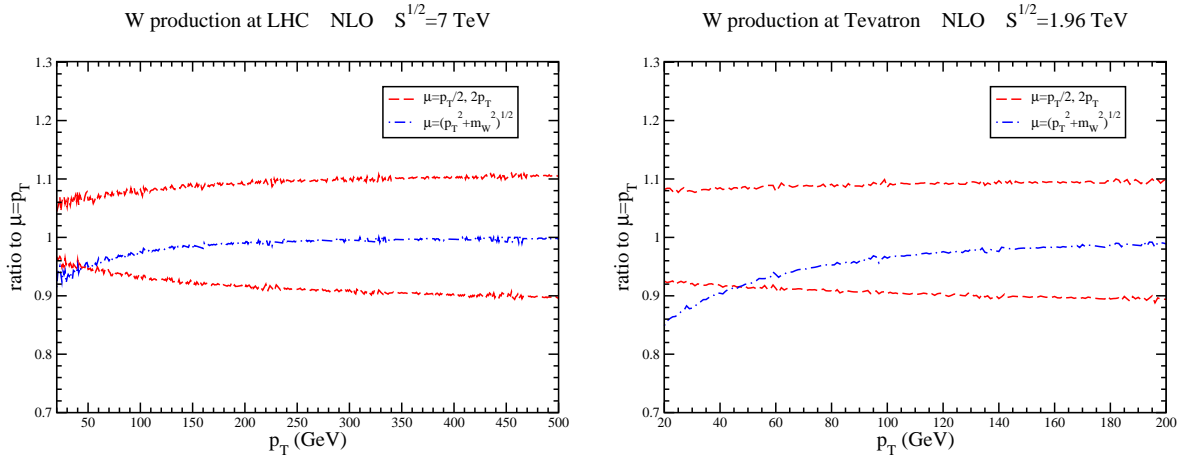

\begin{center}
\includegraphics[width=75mm]{W7lhcNLOmuratioplot.eps}
\hspace{3mm}
\includegraphics[width=75mm]{WtevNLOmuratioplot.eps}
\caption{Ratios of the $W$-boson NLO $p_T$ distribution with various choices of scale to the central result with scale $\mu=p_T$ at the LHC at 7 TeV (left) and at the Tevatron (right).}
\label{Wnlomuratio}
\end{center}
\end{figure}

All numerical results presented
in this paper are for the sum of $W^+$ and $W^-$
differential cross sections.
Initial-state parton densities are taken from MSTW2008 \cite{MSTW}.
We use the NLO central sets with the QCD coupling evolved at NLO in
this section.
In section 4, where we include the NNLL logarithmic corrections, we
employ the NNLO central sets with the QCD coupling evolved at NNLO.
The $W$ is on shell and final-state partons are integrated over the
full phase space.

We begin with results for $W$ production at the LHC at center-of-mass energy
$\sqrt{S} = 7$ TeV and with $p_T$ in the range 20-500 GeV.
At lower values of $p_T$ the fixed order NLO estimates become unreliable due to
Sudakov logarithms and the comparison with experiment needs to include
nonperturbative $p_T$ smearing.
We set the factorization and renormalization scales equal to each other and
denote this common scale by $\mu$.
In the left plot of Fig. \ref{Wnlomuratio} we plot ratios for the $W$-boson $p_T$ distribution 
at the LHC with various choices of scale to the central result with scale $\mu=p_T$.
We display the scale variation of the NLO result with
scale choices $p_T/2$ and $2 p_T$. We also show results for the choice of 
scale $\mu=\sqrt{p_T^2+m_W^2}$ but note that the numbers for this choice 
of scale are very similar to those for $\mu=p_T$. 
We see that the scale variation is of the order of $\pm 10$\%. 
The results for these ratios are almost identical at 8 TeV energy and very similar at 14 TeV.

In the right plot of Fig. \ref{Wnlomuratio} we show results for $W$ production at the
Tevatron at 1.96 TeV energy.
Again, we display the scale variation of the NLO result with
scale choices $p_T/2$ and $2 p_T$ and also $\sqrt{p_T^2+m_W^2}$. 
We note that the published CDF and D0 analyses \cite{CDF,D0} involve
older Run I data.
The results presented here are also applicable to the higher luminosity and
energy data from Run II.

We will say more about the scale variation in Section 4 when we include the NNLO soft-gluon corrections.
Another source of uncertainty in the differential distributions comes from the parton distribution functions
(PDF). We will return to the topic of PDF uncertainties when we present the approximate NNLO results 
in Section 4.3.

\mysection{NNLL Resummation}

Near partonic threshold the corrections from soft-gluon emissions are dominant.
These corrections can be resummed to all orders using renormalization group arguments.
The resummed cross section is derived in Mellin moment space, with $N$
the moment variable conjugate to $s_4$, and is given formally by
\beqa
{\hat{\sigma}}^{res}(N) &=&
\exp\left[ \sum_i E_i(N_i)\right] \, \exp\left[ E'_j(N')\right]\;
\exp \left[\sum_{i} 2 \int_{\mu_F}^{\sqrt{s}} \frac{d\mu}{\mu}\;
\gamma_{i/i}\left({\tilde N}_i, \alpha_s(\mu)\right)\right] \;
\nonumber\\ && \hspace{-20mm} \times \,
H\left(\alpha_s(\sqrt{s})\right) \;
S \left(\alpha_s\left(\frac{\sqrt{s}}{\tilde N'}\right)\right) \;
\exp \left[\int_{\sqrt{s}}^{{\sqrt{s}}/{\tilde N'}}
\frac{d\mu}{\mu}\; 2\, {\rm Re} \Gamma_S\left(\alpha_s(\mu)\right)\right]
\label{sigmares}
\eeqa
where the first exponential resums
the collinear and soft-gluon radiation from the inital-state partons;
the second exponential resums corresponding terms from the final state;
the third exponential controls the factorization scale dependence of the cross section via
the parton-density anomalous dimension;
$H$ is the hard-scattering function; and $S$ is the soft-gluon function describing noncollinear soft gluon
emission whose evolution is controlled by the soft anomalous
dimension $\Gamma_S$. 
The first three exponentials in Eq. (\ref{sigmares}) are independent of
the color structure of the hard scattering and thus universal 
\cite{GS87,CT89}, while the functions $H$, $S$, and $\Gamma_S$ 
are process-specific \cite{NKVD,disdpf11}. 
More details of the resummation formalism
have been given before (see e.g. Refs. \cite{NKVD,disdpf11,NKst})
and will not be repeated here.

\begin{figure}
\begin{center}
\includegraphics[width=10cm]{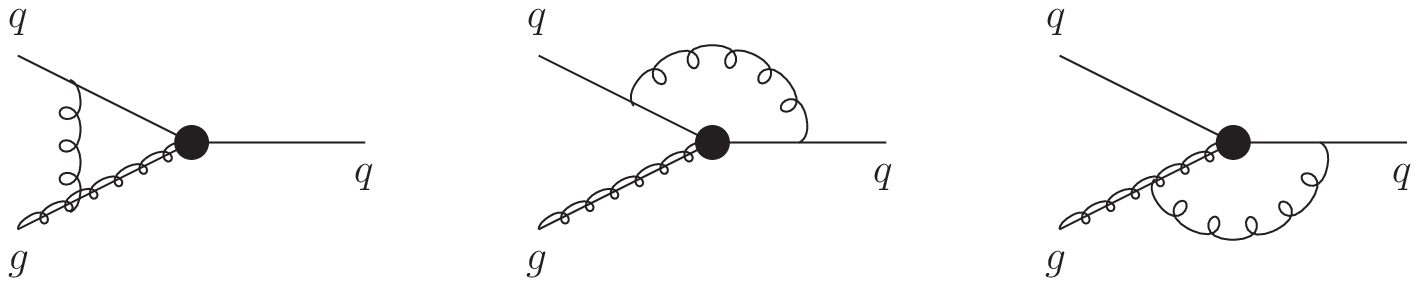}
\caption{One-loop eikonal diagrams for $qg \rightarrow Wq$.}
\label{W1loopqg}
\end{center}
\end{figure}

\begin{figure}
\begin{center}
\includegraphics[width=10cm]{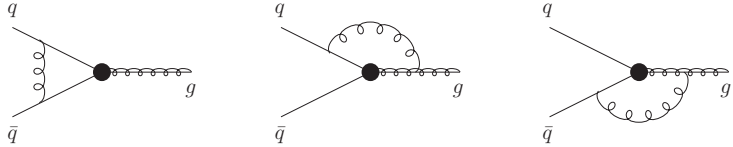}
\caption{One-loop eikonal diagrams for $q{\bar q} \rightarrow Wg$.}
\label{W1loopqq}
\end{center}
\end{figure}

\begin{figure}
\begin{center}
\includegraphics[width=70mm]{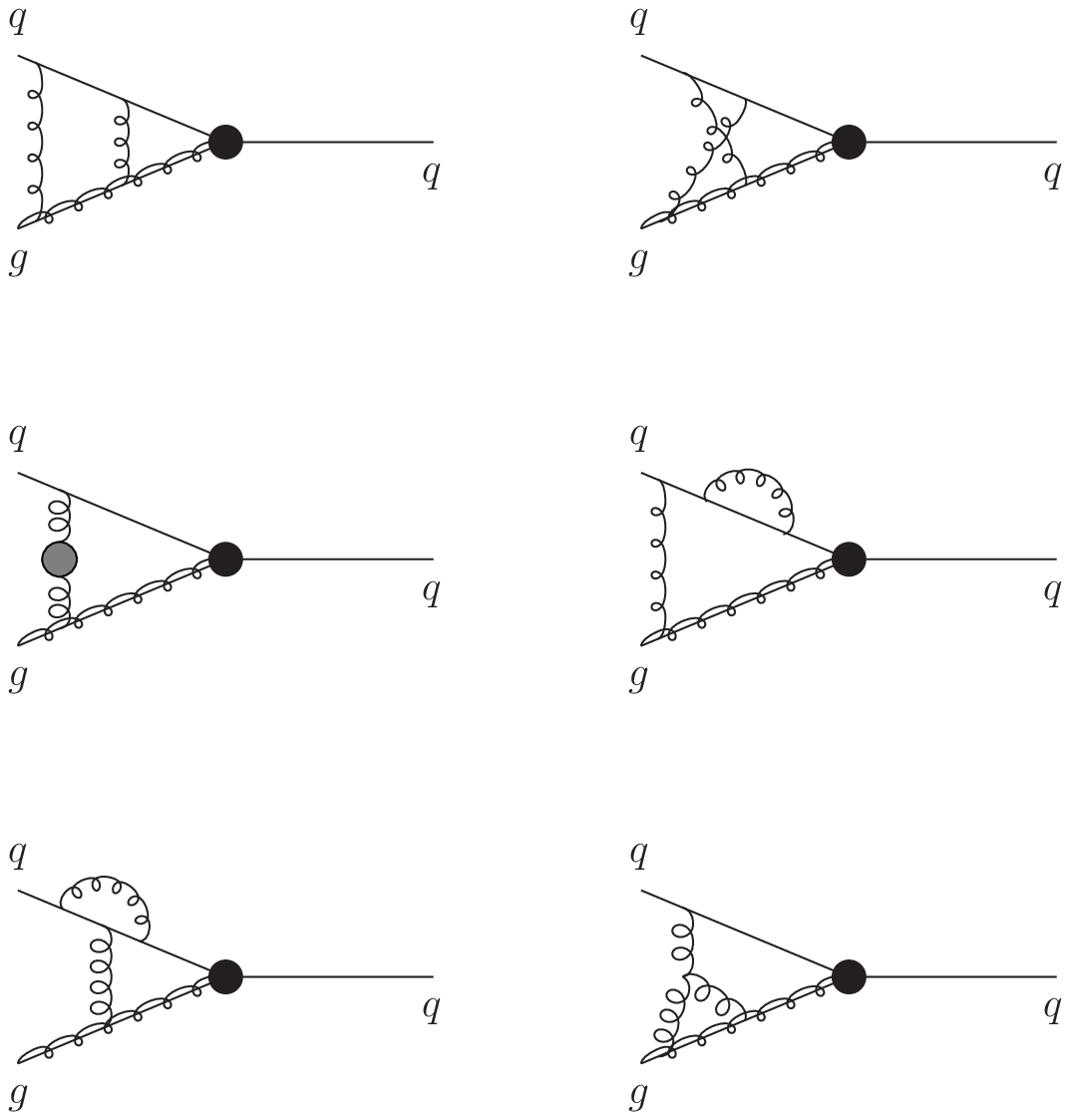}
\hspace{15mm}
\includegraphics[width=70mm]{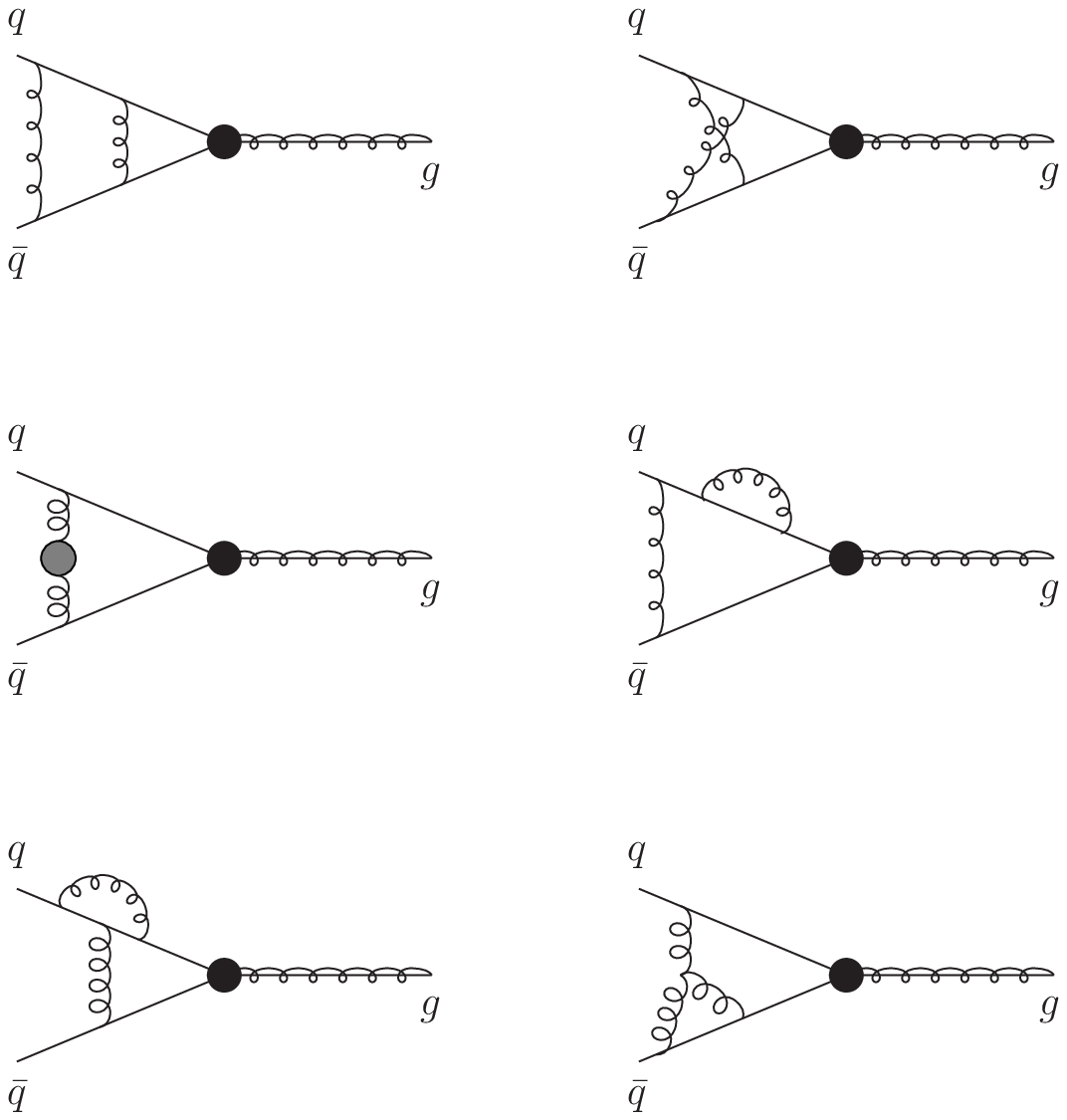}
\caption{Two-loop eikonal diagrams for $qg \rightarrow Wq$ and $q{\bar q} \rightarrow Wg$ involving the two incoming partons. 
There are two additional sets
of 12 diagrams each with the same topologies that involve one incoming and
one outgoing parton.}
\label{W2loop1}
\end{center}
\end{figure}

We expand the process-specific soft anomalous dimensions $\Gamma_S$
in the strong coupling as
\beq
\Gamma_S=\frac{\alpha_s}{\pi}\Gamma_S^{(1)}
+\frac{\alpha_s^2}{\pi^2}\Gamma_S^{(2)}
+\cdots \, .
\eeq
The one-loop results, $\Gamma_S^{(1)}$, are obtained from the ultraviolet poles in dimensional
regularization of one-loop eikonal diagrams involving the colored particles in the partonic processes,
Figs. \ref{W1loopqg} and \ref{W1loopqq},
and were first derived in \cite{NKVD}.
We determine the two-loop results, $\Gamma_S^{(2)}$, from the ultraviolet poles of two-loop
dimensionally regularized integrals for eikonal diagrams shown in
Fig. \ref{W2loop1} and related graphs involving other combinations of the eikonal lines (see also \cite{NKRGdpf}).

For $qg\rightarrow Wq$ the one-loop soft anomalous dimension is
\beq
\Gamma_{S,\, qg\rightarrow Wq}^{(1)}=C_F \ln\left(\frac{-u}{s}\right)
+\frac{C_A}{2} \ln\left(\frac{t}{u}\right)
\eeq
and the two-loop soft anomalous dimension is
\beq
\Gamma_{S,\, qg \rightarrow Wq}^{(2)}=\frac{K}{2} \Gamma_{S,\, qg \rightarrow Wq
}^{(1)} \, ,
\eeq
where $K=C_A(67/18-\zeta_2)-5n_f/9$ \cite{KT82} 
with $C_A=3$ and $n_f$ the number of
light quark flavors.

For $q {\bar q}\rightarrow Wg$ the corresponding results are
\beq
\Gamma_{S,\, q{\bar q}\rightarrow Wg}^{(1)}=\frac{C_A}{2} \ln\left(\frac{tu}{s^2
}\right)
\eeq
and
\beq
\Gamma_{S,\, q{\bar q} \rightarrow Wg}^{(2)}=\frac{K}{2} \Gamma_{S,\, q{\bar q}
\rightarrow Wg}^{(1)} \, .
\eeq

We note that the proportionality of the two-loop soft anomalous dimension to the one-loop result is 
anticipated on general grounds from the work in Ref. \cite{ADS} (see also \cite{DMS,BN,GM}).

\mysection{NNLO approximate results}

By expanding the resummed cross section, Eq. (\ref{sigmares}), in the strong coupling
we derive approximate fixed-order results. In this section we present the analytical expressions for the NNLO expansion and use them to present approximate NNLO
results for the $W$-boson transverse
momentum distribution at the LHC and the Tevatron.

\subsection{$qg \longrightarrow Wq$}

We can write the NLO soft and virtual corrections for
$qg \rightarrow Wq$ as
\beq
E_Q\frac{d{\hat\sigma}^{(1)}_{qg \rightarrow Wq}}{d^3Q} =
F^B_{qg \rightarrow Wq}
{\alpha_s(\mu_R^2)\over\pi}\,
\left\{c_3^{qg} \, \left[\frac{\ln(s_4/p_T^2)}{s_4}\right]_+
+c_2^{qg} \, \left[\frac{1}{s_4}\right]_+ + c_1^{qg} \,
\delta(s_4)\right\} \, .
\label{qgnlo}
\eeq
The NLO coefficients $c_3^{qg}$ and $c_2^{qg}$ of the soft-gluon terms
in Eq. (\ref{qgnlo}) can be derived from the expansion of the resummed cross
section and are given by
$c_3^{qg}=C_F+2C_A$ and
\beq
c_2^{qg}=-\left(C_F + C_A\right) \ln\left(\frac{\mu_F^2}{p_T^2}\right)
- \frac{3}{4} C_F - C_A \ln{\left(\frac{t u}{s p_T^2}\right)} \, .
\eeq
For later use in the NNLO results we also define the scale-independent
part of $c_2^{qg}$ as $T_2^{qg}=-(3/4) C_F - C_A \ln(t u/s p_T^2)$.

The coefficient of the $\delta(s_4)$ terms in Eq. (\ref{qgnlo}) is given by
\beq
c_1^{qg}=\frac{1}{2A^{qg}}\left[B_1^{qg}+B_2^{qg} n_f
+C_1^{qg}+C_2^{qg} n_f \right]+\frac{c_3^{qg}}{2}
\ln^2\left(\frac{p_T^2}{Q^2}\right)
+c_2^{qg} \ln\left(\frac{p_T^2}{Q^2}\right)\, ,
\eeq
with $B_1^{qg}$, $B_2^{qg}$, $C_1^{qg}$, and $C_2^{qg}$
as given in the Appendix of the first paper in Ref. \cite{gpw} 
but without the renormalization counterterms
and using $f_A \equiv\ln(A/Q^2)=0$ [note that the terms not multiplying
$A^{qg}$ in Eq. (A4) for $B_1^{qg}$ of Ref. \cite{gpw} should have the
opposite sign than shown in that paper].

The NNLO expansion of the resummed cross section  involves logarithms $\ln^k(s_4/p_T^2)$ with $k=0,1,2,3$.
The terms with $k=1,2,3$ were already provided in
Refs. \cite{NKVD,NKASV} from NLL resummation.
Terms for $k=0$ were also given in Eq. (3.8) of Ref. \cite{NKASV} but
they were incomplete. Now that we have expressions for NNLL resummation we can provide
the complete result for the $k=0$ terms.

The NNLO soft-gluon corrections for
$qg \rightarrow Wq$ can be written as
\beq
E_Q\frac{d{\hat\sigma}^{(2)}_{qg \rightarrow Wq}}{d^3Q} =
F^B_{qg \rightarrow Wq}
\frac{\alpha_s^2(\mu_R^2)}{\pi^2} \,
{\hat{\sigma'}}^{(2)}_{qg \rightarrow Wq}
\label{NNLOmqg}
\eeq
with
\beqa
{\hat{\sigma'}}^{(2)}_{qg \rightarrow Wq}&=&
\frac{1}{2} (c_3^{qg})^2 \, \left[\frac{\ln^3(s_4/p_T^2)}{s_4}\right]_+
+\left[\frac{3}{2} c_3^{qg} \, c_2^{qg}
- \frac{\beta_0}{4} c_3^{qg}
+C_F \frac{\beta_0}{8}\right] \left[\frac{\ln^2(s_4/p_T^2)}{s_4}\right]_+
\nonumber \\ && \hspace{-5mm}
{}+\left\{c_3^{qg} \, c_1^{qg} +(c_2^{qg})^2
-\zeta_2 \, (c_3^{qg})^2 -\frac{\beta_0}{2} \, T_2^{qg}
+\frac{\beta_0}{4} c_3^{qg} \ln\left(\frac{\mu_R^2}{p_T^2}\right) \right.
\nonumber \\ && \hspace{-5mm} \quad  \left.
{}+(C_F+2C_A)\frac{K}{2}
-\frac{3}{16} \beta_0 C_F \right\}
\left[\frac{\ln(s_4/p_T^2)}{s_4}\right]_+
\nonumber \\ && \hspace{-5mm}
{}+\left\{c_2^{qg} \, c_1^{qg}
-\zeta_2 \, c_3^{qg} \, c_2^{qg}
+\zeta_3 \, (c_3^{qg})^2
+\frac{\beta_0}{4}\, c_2^{qg} \ln\left(\frac{\mu_R^2}{s}\right)
-\frac{\beta_0}{2} C_F \ln^2\left(\frac{-u}{p_T^2}\right)\right.
\nonumber \\ && \hspace{-5mm} \quad
{}-\frac{\beta_0}{2} C_A \ln^2\left(\frac{-t}{p_T^2}\right)
-C_F\,  K \,  \ln\left(\frac{-u}{p_T^2}\right)
-C_A\,  K \,  \ln\left(\frac{-t}{p_T^2}\right)
\nonumber \\ && \hspace{-5mm} \quad
{}+(C_F+C_A) \left[\frac{\beta_0}{8}
\ln^2\left(\frac{\mu_F^2}{s}\right)
- \frac{K}{2} \, \ln\left(\frac{\mu_F^2}{s}\right)\right]
-\left(C_F \frac{K}{2}-\frac{3\beta_0 C_F}{16}\right)
\ln\left(\frac{p_T^2}{s}\right)
\nonumber \\ &&  \hspace{-5mm} \quad \left.
{}+\frac{3 \beta_0}{8} C_F \ln^2\left(\frac{p_T^2}{s}\right)
+2 \, D_q^{(2)} +D_g^{(2)} +B_q^{(2)}
+2\, \Gamma^{(2)}_{S\, qg\rightarrow W q} \right\} \,
\left[\frac{1}{s_4}\right]_+  \,
\eeqa
with $\beta_0=(11 C_A-2 n_f)/3$ and
where we have used the two-loop constants (cf. \cite{CLS97,MVV02})
\beq
D_q^{(2)}=C_F C_A \left(-\frac{101}{54}+\frac{11}{6} \zeta_2
+\frac{7}{4}\zeta_3\right)
+C_F n_f \left(\frac{7}{27}-\frac{\zeta_2}{3}\right) \, ,
\eeq
$D_g^{(2)}=(C_A/C_F) D_q^{(2)}$, and
\beq
B_q^{(2)}=C_F^2\left(-\frac{3}{32}+\frac{3}{4}\zeta_2-\frac{3}{2}\zeta_3\right)
+C_F C_A \left(-\frac{1539}{864}-\frac{11}{12}\zeta_2+\frac{3}{4}\zeta_3\right)
+n_f C_F \left(\frac{135}{432}+\frac{\zeta_2}{6}\right) \, .
\eeq

Note that the difference from Eq. (3.8) of Ref. \cite{NKASV} is in the
$[1/s_4]_+$ terms. The higher powers of the logarithms are the same.
Also note that one can determine the scale-dependent $\delta(s_4)$ terms
at NNLO. These terms are also given in Ref. \cite{NKASV}
and will not be repeated here.

\subsection{$q{\bar q} \longrightarrow Wg$}

For the process $q{\bar q} \rightarrow Wg$
we can write the NLO soft and virtual corrections as
\beqa
E_Q\frac{d{\hat\sigma}^{(1)}_{q{\bar q} \rightarrow Wg}}{d^3Q} &=&
{F^B_{q{\bar q} \rightarrow Wg}} {\alpha_s(\mu_R^2)\over\pi}\,
\left\{c_3^{q \bar q} \, \left[\frac{\ln(s_4/p_T^2)}{s_4}\right]_+
+c_2^{q \bar q} \, \left[\frac{1}{s_4}\right]_+
+c_1^{q \bar q} \, \delta(s_4)\right\} \, .
\label{qqbarnlo}
\eeqa
Here the NLO coefficients $c_3^{q{\bar q}}$ and $c_2^{q{\bar q}}$ of the soft-gluon terms
are $c_3^{q \bar q}=4C_F-C_A$ and
\beq
c_2^{q \bar q}=- 2 C_F \ln\left(\frac{\mu_F^2}{p_T^2}\right)
- \left(2 C_F- C_A \right) \ln\left(\frac{t u}{s p_T^2}\right)
-\frac{\beta_0}{4} \, .
\eeq
For later use in the NNLO results we also define the scale-independent
part of $c_2^{q{\bar q}}$ as
$T_2^{q{\bar q}}=-(2 C_F- C_A) \ln(t u/s p_T^2) -\beta_0/4$.

The coefficient of the $\delta(s_4)$ terms in Eq. (\ref{qqbarnlo}) is given by
\beq
c_1^{q \bar q}=\frac{1}{2A^{q \bar q}}\left[B_1^{q \bar q}+C_1^{q \bar q}
+(B_2^{q \bar q}+D_{aa}^{(0)}) \, n_f \right]
+\frac{c_3^{q \bar q}}{2} \ln^2\left(\frac{p_T^2}{Q^2}\right)
+c_2^{q \bar q} \ln\left(\frac{p_T^2}{Q^2}\right)\, ,
\eeq
with $B_1^{q \bar q}$, $B_2^{q \bar q}$,
$C_1^{q \bar q}$, and $D_{aa}^{(0)}$
as given in the Appendix of Ref. \cite{gpw} but without the renormalization
counterterms and using $f_A=0$.

Again, the NNLO expansion involves logarithms $\ln^k(s_4/p_T^2)$
with $k=0,1,2,3$. The terms with $k=1,2,3$ were already provided in
Refs. \cite{NKVD,NKASV} from NLL resummation, but the terms given
for $k=0$ in Eq. (3.19) in \cite{NKASV} were incomplete.
With NNLL resummation we can now provide
the complete result.

The NNLO soft-gluon corrections for
$q{\bar q} \rightarrow Wg$ can be written as
\beq
E_Q\frac{d{\hat\sigma}^{(2)}_{q{\bar q} \rightarrow Wg}}{d^3Q} =
{F^B_{q{\bar q} \rightarrow Wg}}
\frac{\alpha_s^2(\mu_R^2)}{\pi^2} \, {\hat{\sigma'}}^{(2)}_{q{\bar q}
\rightarrow Wg}
\label{NNLOmqqb}
\eeq
with
\beqa
{\hat{\sigma'}}^{(2)}_{q{\bar q} \rightarrow Wg}&=&
\frac{1}{2} (c_3^{q \bar q})^2 \, \left[\frac{\ln^3(s_4/p_T^2)}{s_4}\right]_+
+\left[\frac{3}{2} c_3^{q \bar q} \, c_2^{q \bar q}
- \frac{\beta_0}{4} c_3^{q \bar q}
+C_A \frac{\beta_0}{8}\right] \left[\frac{\ln^2(s_4/p_T^2)}{s_4}\right]_+
\nonumber \\ && \hspace{-5mm}
{}+\left\{c_3^{q \bar q} \, c_1^{q \bar q} +(c_2^{q \bar q})^2
-\zeta_2 \, (c_3^{q \bar q})^2 -\frac{\beta_0}{2} \, T_2^{q \bar q}
+\frac{\beta_0}{4} c_3^{q \bar q}  \ln\left(\frac{\mu_R^2}{p_T^2}\right)
\right.
\nonumber \\ && \hspace{-5mm} \quad \left.
{}+(4C_F-C_A) \frac{K}{2}
-\frac{\beta_0^2}{16} \right\}
\left[\frac{\ln(s_4/p_T^2)}{s_4}\right]_+
\nonumber \\ && \hspace{-5mm}
{}+\left\{c_2^{q{\bar q}} \, c_1^{q{\bar q}}
-\zeta_2 \, c_3^{q{\bar q}} \, c_2^{q{\bar q}}
+\zeta_3 \, (c_3^{q{\bar q}})^2
+\frac{\beta_0}{4}\, c_2^{q{\bar q}} \ln\left(\frac{\mu_R^2}{s}\right)
-\frac{\beta_0}{2} C_F \ln^2\left(\frac{-u}{p_T^2}\right)\right.
\nonumber \\ && \hspace{-5mm} \quad
{}-\frac{\beta_0}{2} C_F \ln^2\left(\frac{-t}{p_T^2}\right)
+C_F \left[-K \,  \ln\left(\frac{tu}{p_T^4}\right)
+\frac{\beta_0}{4} \ln^2\left(\frac{\mu_F^2}{s}\right)
- K \, \ln\left(\frac{\mu_F^2}{s}\right)\right]
\nonumber \\ && \hspace{-5mm} \quad
{}-\left(C_A \frac{K}{2}-\frac{\beta_0^2}{16}\right)
\ln\left(\frac{p_T^2}{s}\right)
+\frac{3 \beta_0}{8} C_A \ln^2\left(\frac{p_T^2}{s}\right)
\nonumber \\ && \hspace{-5mm} \quad \left.
{}+2 \, D_q^{(2)} +D_g^{(2)} +B_g^{(2)}
+2 \, \Gamma^{(2)}_{S\, q{\bar q}\rightarrow W g} \right\} \,
\left[\frac{1}{s_4}\right]_+
\eeqa
where
\beq
B_g^{(2)}=C_A^2\left(-\frac{1025}{432}-\frac{3}{4}\zeta_3\right)
+\frac{79}{108} C_A \, n_f +C_F \frac{n_f}{8}-\frac{5}{108} n_f^2 \, .
\eeq

The scale-dependent $\delta(s_4)$ terms at NNLO were also provided in
Eq. (3.19) of \cite{NKASV} and we will not repeat them here.

\subsection{Numerical results}

\begin{figure}
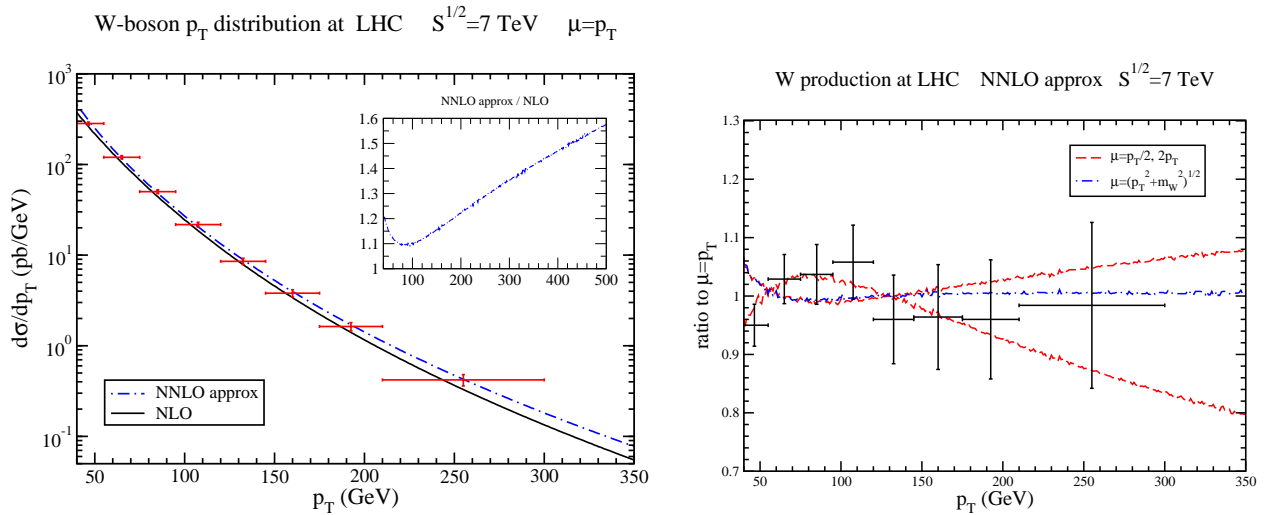

\begin{center}
\includegraphics[width=85mm]{W7lhcnnloratiodsdptplot.eps}
\hspace{3mm}
\includegraphics[width=75mm]{W7lhcnnlomuratioplot.eps}
\caption{$W$-boson approximate NNLO $p_T$ distribution at the LHC at 7 TeV compared with ATLAS data.}
\label{Wlhc7nnlo}
\end{center}
\end{figure}

We begin with results for $W$ production at the LHC at 7 TeV energy.
In Fig. \ref{Wlhc7nnlo} we plot the $W$-boson $p_T$ distribution, 
$d\sigma/dp_T$. 
In the left plot, we compare the NLO and the approximate NNLO results at the LHC at 7 TeV energy 
with $\mu=p_T$. We also compare our results to recent data from ATLAS 
\cite{ATLAS}. It is evident that the effect of the NNLO soft-gluon corrections 
grows with $p_T$ as one would expect,
since the kinematical region near partonic threshold becomes more important
at higher $p_T$. 
The inset plot shows that the ratio of the approximate NNLO 
to the full NLO (i.e. the $K$ factor) grows with $p_T$, and the NNLO 
soft-gluon corrections provide nearly a 60\% enhancement at $p_T=500$ GeV.
Since the ATLAS data use acceptance cuts and are normalized by the total 
fiducial cross section, we have to correct for these factors to extrapolate 
the experimental results for direct comparison to our $p_T$ distribution.
We use the procedure described in \cite{BLS2}. We multiply the normalized 
ATLAS results by the total fiducial cross section and divide by the acceptances.
It is clear from the 
comparison that the data are in very good agreement with our NNLO approximate 
result, which provides a better description than NLO alone. The ATLAS data 
only go up to a $p_T$ of 300 GeV and it will be interesting to 
see data from the LHC at even higher $p_T$. 

In the right plot of Fig. \ref{Wlhc7nnlo} we show ratios of the approximate NNLO result with the variation of the 
result with scale $\mu=p_T/2$ and $2 p_T$ relative to $\mu=p_T$. We also show the ratio  
with $\mu=\sqrt{p_T^2+m_W^2}$ and note that the results for this choice 
of scale are very similar to those for $\mu=p_T$. 
Also, comparing the left plot of Fig. \ref{Wnlomuratio} with 
the right plot of Fig. \ref{Wlhc7nnlo}
it is seen that the scale dependence at approximate NNLO is smaller than at NLO
at intermediate values of $p_T$, but it grows larger at very high $p_T$ values
where the overall soft-gluon contribution is also larger.

We note that results for 7 TeV LHC energy have also been provided in 
Ref. \cite{BLS} at NNLL in the SCET formalism. It is not possible to do an 
analytical comparison with \cite{BLS} because no analytical results are 
provided there. It is also important to note that NNLL means different things in different approaches and also if one uses different variables
within the same approach. This has been clearly explained in the context of top
quark production in the review paper of Ref. \cite{NKBP} and applies here as well. 
Therefore our NNLL Mellin-space resummation is not equivalent to the NNLL SCET resummation of \cite{BLS}.
Related studies for top quark cross sections show that different formalisms can give very different numerical results \cite{NKBP}.

Furthermore, it is difficult to make a meaningful numerical comparison  
because most of the figures in \cite{BLS} plot results versus parameters that 
are only defined within the SCET formalism. Non-graphical results are 
provided in Table 1 of \cite{BLS}, which provides absolute integrated
cross section estimates for $p_T > 200$ GeV. The LO and NLO fixed-order results
are of course in agreement with ours, but exponentiated
logarithmic corrections will obviously diverge in this bin, thus making the
comparison not very meaningful. At a purely numerical level, however, we note that our approximate NNLO integrated cross section at 7 TeV LHC energy
above a $p_T$ of 200 GeV is about 20\% larger than that provided in Table 1 of Ref. \cite{BLS}. 

Finally it is important to note that here we use an NNLO expansion of the resummed
expression to obtain numerical results, in order to avoid prescriptions needed
to invert from moment space to momentum space. Even within a given formalism
such as SCET or Mellin moments, there is a numerical difference between using
NNLO expansions and attempting a full resummation (again see \cite{NKBP}  
and references therein for more details). However, this difference is typically smaller 
than the differences between different formalisms or prescriptions, so this is usually a point 
of relatively minor significance. For example both the values and uncertainties in \cite{BLS} 
are very similar for the resummed and NNLO approximate results.

\begin{figure}
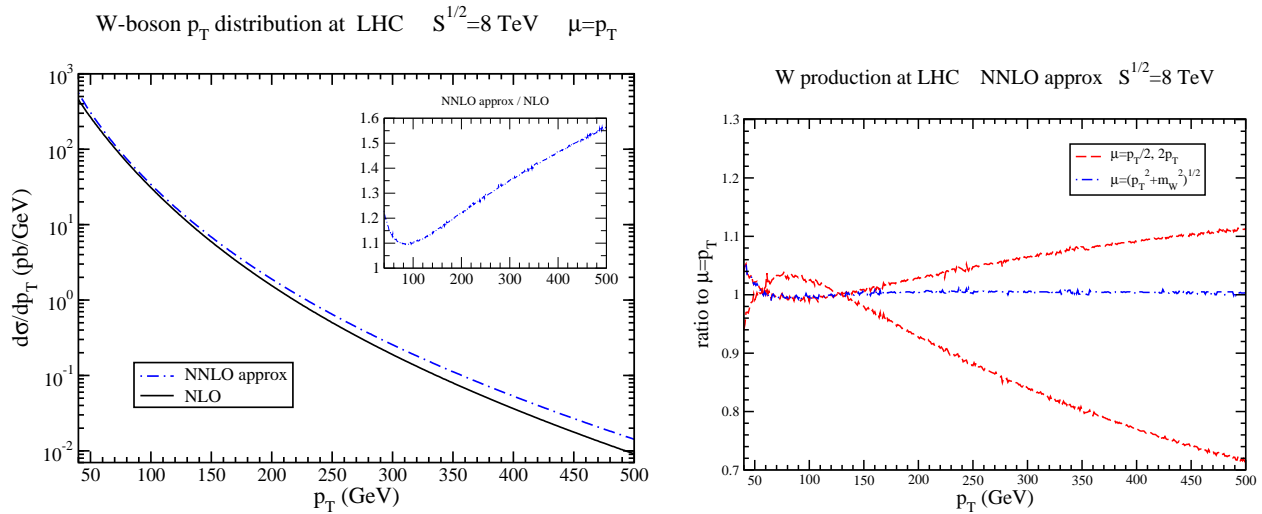

\begin{center}
\includegraphics[width=85mm]{W8lhcnnloratiodsdptplot.eps}
\hspace{3mm}
\includegraphics[width=75mm]{W8lhcnnlomuratioplot.eps}
\caption{$W$-boson approximate NNLO $p_T$ distribution at the LHC at 8 TeV.}
\label{Wlhc8nnlo}
\end{center}
\end{figure}

We continue with the presentation of our numerical results for the $W$-boson $p_T$ distribution at the LHC. 
In Fig. \ref{Wlhc8nnlo} we show results for the LHC at 8 TeV energy. 
Although the overall $p_T$ distribution is enhanced 
at 8 TeV relative to 7 TeV, the result for the ratio of approximate NNLO over NLO shown in the inset 
plot is very similar (almost identical) to that at 7 TeV. Also the scale dependence at 8 TeV is almost the 
same as at 7 TeV as can be seen by comparing the right plots of Figs. \ref{Wlhc7nnlo} and \ref{Wlhc8nnlo}.

\begin{figure}
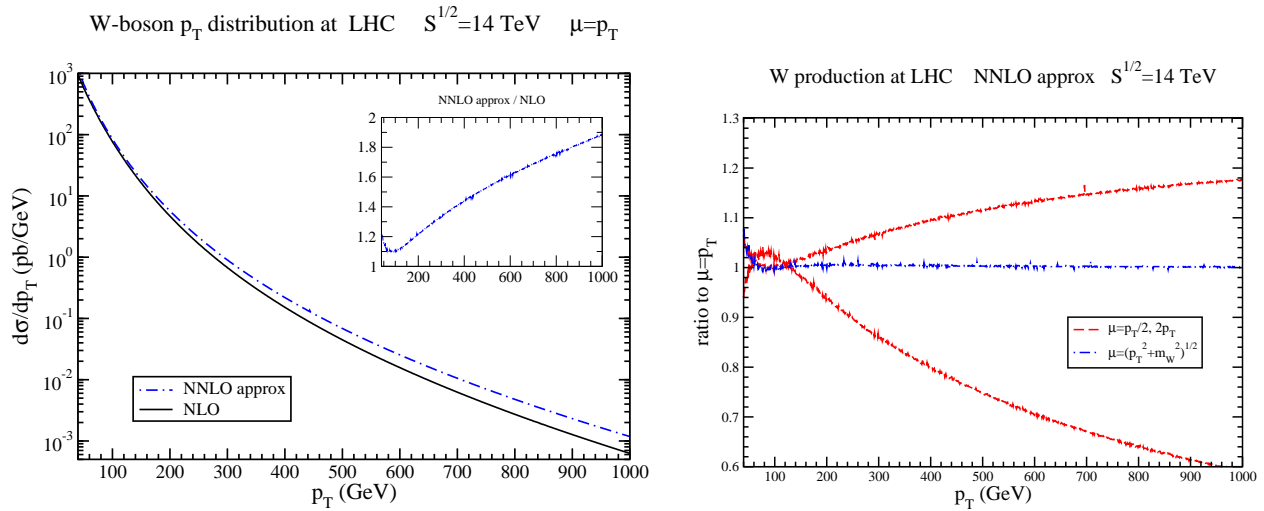

\begin{center}
\includegraphics[width=85mm]{W14lhcnnloratiodsdptplot.eps}
\hspace{3mm}
\includegraphics[width=75mm]{W14lhcnnlomuratioplot.eps}
\caption{$W$-boson approximate NNLO $p_T$ distribution at the LHC at 14 TeV.}
\label{Wlhc14nnlo}
\end{center}
\end{figure}

In Fig. \ref{Wlhc14nnlo} we show the corresponding results for the LHC at 14 TeV energy.
Again, the increase due to the NNLO soft-gluon corrections is more prominent at very high $p_T$, 
reaching 90\% enhancement at $p_T=1000$ GeV.
The scale variation at the LHC at 14 TeV energy is shown on the right plot. 
At the very highest $p_T$ the scale uncertainty is significant. 

\begin{figure}
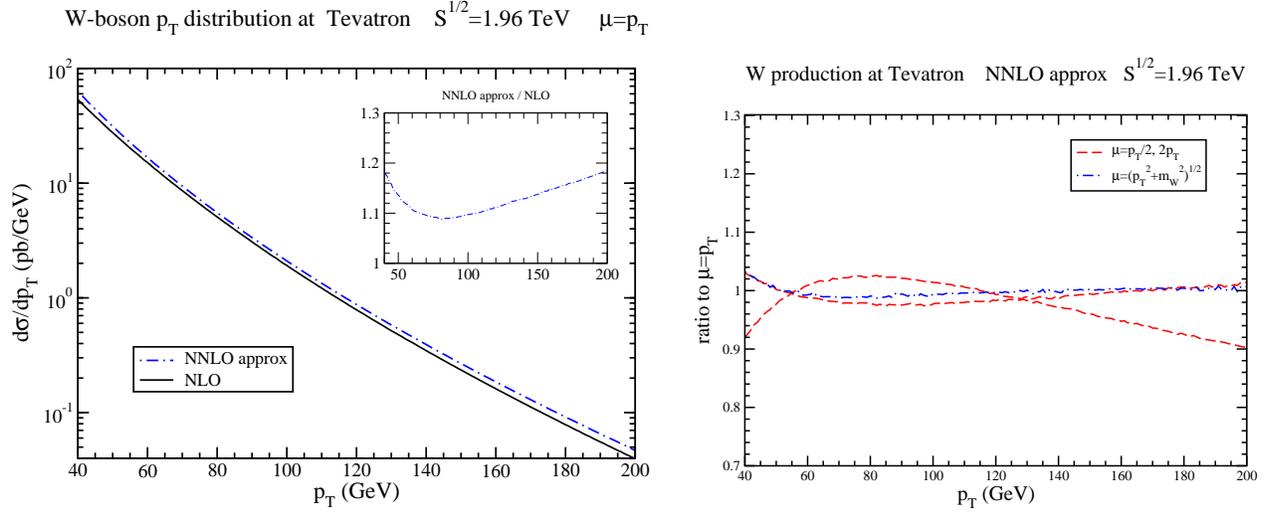

\begin{center}
\includegraphics[width=85mm]{Wtevnnloratiodsdptplot.eps}
\hspace{3mm}
\includegraphics[width=75mm]{Wtevnnlomuratioplot.eps}
\caption{$W$-boson approximate NNLO $p_T$ distribution at the Tevatron.}
\label{Wtevnnlo}
\end{center}
\end{figure}

Finally, we present results for the Tevatron. The left plot of Fig. \ref{Wtevnnlo} shows the NLO and NNLO approximate $p_T$ distributions at the Tevatron at $\mu=p_T$, 
with the inset plot displaying the $K$ factor. 
The right plot of Fig. \ref{Wtevnnlo} shows the results for scale variation at the Tevatron, where there is a 
reduction of scale dependence at NNLO relative to that at NLO in the $p_T$ range shown, as can be seen by comparing the right plots of Figs. \ref{Wnlomuratio} and \ref{Wtevnnlo}.

In addition to the scale variation shown in the previous figures, one can also attempt an 
independent variation of the factorization and renormalization scales; however,
this does not affect the range of the overall uncertainty significantly, if at all.

\begin{figure}
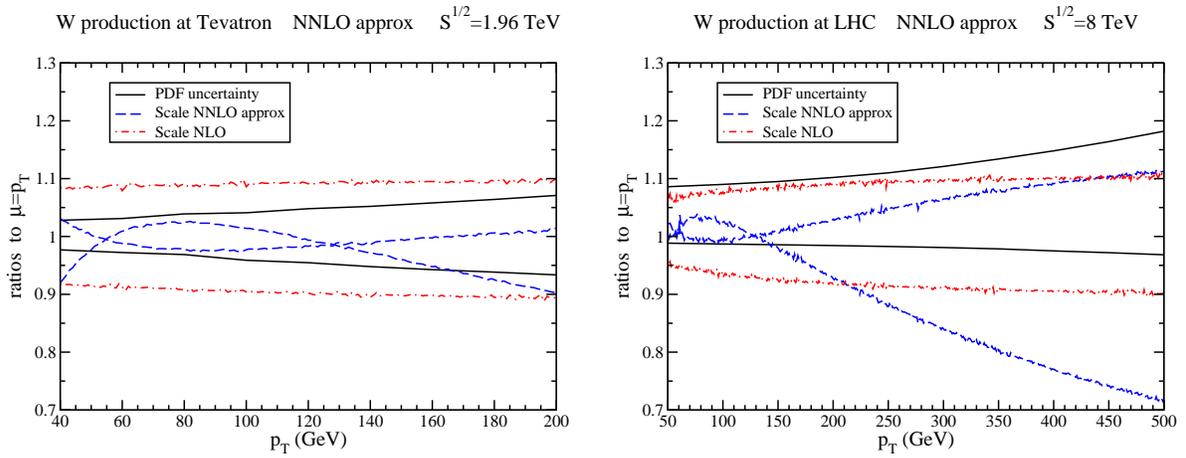

\begin{center}
\includegraphics[width=75mm]{pdftevplot.eps}
\hspace{3mm}
\includegraphics[width=75mm]{pdf8lhcplot.eps}
\caption{PDF and scale uncertainties for the $W$-boson $p_T$ distribution at the Tevatron (left) and at the LHC at 
8 TeV (right).}
\label{Wpdf}
\end{center}
\end{figure}

As already mentioned in Section 2, in addition to scale dependence another important source of uncertainty 
comes from the PDF. Here we use the PDF sets and procedure provided by MSTW2008 \cite{MSTW} to calculate the PDF uncertainties for $W$ production.
In the left plot of Fig. \ref{Wpdf} we compare the PDF uncertainty with the scale uncertainty at NNLO and also 
at NLO at the Tevatron. The scale ratios are for $\mu=p_T/2$ and $2p_T$ and are the same as already shown in the plots  
of the previous Tevatron figures, but displaying them together and with the PDF ratios makes the 
comparison of all the uncertainties easier. We note that for the Tevatron the PDF uncertainties are smaller than the NLO scale 
variation, but they are larger than the scale variation at approximate NNLO for most $p_T$ values. As mentioned earlier, 
the NNLO scale variation is consistently smaller than that at NLO.

In the right plot of Fig. \ref{Wpdf} we compare the PDF uncertainty with the scale uncertainty at NNLO and 
also at NLO at the LHC at 8 TeV energy (the results at 7 TeV are practically the same). Here the situation 
is somewhat different from the Tevatron in that the upper range of the PDF uncertainty is larger 
than both NLO and NNLO scale variation, but the lower range is smaller than both 
for most $p_T$ values. Also the scale variation at approximate NNLO is smaller than that at NLO 
for moderate $p_T$ values but becomes larger, particularly at the lower end, at very high $p_T$.

The growing NNLO contribution relative to NLO with increasing
$p_T$ at both Tevatron and LHC energies
does not appear to be an artifact of perturbation theory.
The extended $p_T$ range made accessible by the LHC brings into play a high
multiplicity of hard jets in various inclusive cross sections, and this is
expected to affect the convergence of perturbative predictions.
The recent ATLAS data \cite{ATLAS} and its agreement with our theoretical 
results confirms this expectation. 
An analysis of the $Z$ distribution is underway and will be reported elsewhere.

\mysection{Conclusions}

The transverse momentum distribution of the $W$ boson receives large
QCD corrections. Complete NLO calculations have been used in this paper
to provide numerical results at LHC and Tevatron energies.
In addition, NNLL resummation of soft-gluon corrections has been derived
using two-loop soft anomalous dimensions. Approximate NNLO analytical expressions
have been derived from the resummed cross section and employed to produce numerical results.
The NNLO soft-gluon corrections reduce the NLO scale dependence at low and intermediate $p_T$
where the bulk of the data is located.  In the very high $p_T$ region
at LHC energies the soft logarithms seem to become sensitive to
scale. It is possible that including hard NNLO corrections will improve this situation.
The experimental bins at large $p_T$ are larger because the number of expected events
and the $p_T$ resolution both decrease dramatically. Recent ATLAS data 
\cite{ATLAS} are in good agreement with our numerical results. 
The higher-order results presented in this paper strengthen theoretical
predictions for $W$ production at hadron colliders and may prove significant
for new physics searches in the tail of the $p_T$ spectrum.

\mysection*{Acknowledgements}
The work of N.K. was supported by the National Science Foundation under
Grant No. PHY 0855421.


\begin{thebibliography}{99}

\bibitem{AR}
P.B. Arnold and M.H. Reno,  Nucl. Phys. B {\bf 319}, 37 (1989);
(E) B {\bf 330}, 284 (1990).

\bibitem{gpw}
R.J. Gonsalves, J. Paw{\l}owski, and C.-F. Wai, Phys. Rev. D {\bf 40},
2245 (1989); Phys. Lett. B {\bf 252}, 663 (1990).

\bibitem{CDF}
CDF Collaboration, F. Abe {\it et al.}, Phys. Rev. Lett. {\bf 66}, 2951 (1991).

\bibitem{D0}
D0 Collaboration, B. Abbott {\it et al.}, Phys. Rev. Lett. {\bf 80}, 5498 (1998) [hep-ex/9803003];
V.M. Abazov {\it et al.}, Phys. Lett. B {\bf 513}, 292 (2001) [hep-ex/0010026].

\bibitem{NKVD}
N. Kidonakis and V. Del Duca, Phys. Lett. B {\bf 480}, 87 (2000) [hep-ph/9911460].

\bibitem{NKASV}
N. Kidonakis and A. Sabio Vera, JHEP {\bf 02}, 027 (2004) [hep-ph/0311266].

\bibitem{GKS}
R.J. Gonsalves, N. Kidonakis, and A. Sabio Vera, Phys. Rev. Lett. {\bf 95},
222001 (2005) [hep-ph/0507317].

\bibitem{NKRGdpf}
N. Kidonakis and R.J. Gonsalves, in {\sl DPF 2011}, arXiv:1109.2817 [hep-ph].

\bibitem{BLS}
T. Becher, C. Lorentzen, and M.D. Schwartz, Phys. Rev. Lett. {\bf 108}, 012001 (2012) [arXiv:1106.4310 [hep-ph]].

\bibitem{ATLAS}
ATLAS Collaboration, G. Aad {\it et al.}, Phys. Rev. D {\bf 85}, 012005 (2012) [arXiv:1108.6308 [hep-ex]].

\bibitem{MSTW}
A.D. Martin, W.J. Stirling, R.S. Thorne, and G. Watt,
Eur. Phys. J. C {\bf 63}, 189 (2009) [arXiv:0901.0002 [hep-ph]].

\bibitem{GS87}
G. Sterman, Nucl. Phys. B {\bf 281}, 310 (1987).

\bibitem{CT89}
S. Catani and L. Trentadue, Nucl. Phys. B {\bf 327}, 323 (1989).

\bibitem{disdpf11}
N. Kidonakis, in {\sl DIS 2011}, arXiv:1105.4267 [hep-ph];
in {\sl DPF 2011}, arXiv:1109.1578 [hep-ph].

\bibitem{NKst}
N. Kidonakis, Phys. Rev. D {\bf 81}, 054028 (2010) [arXiv:1001.5034 [hep-ph]];
Phys. Rev. D {\bf 82}, 054018 (2010) [arXiv:1005.4451 [hep-ph]];
Phys.Rev. D {\bf 83}, 091503 (2011) [arXiv:1103.2792 [hep-ph]].

\bibitem{KT82} K
J. Kodaira and L. Trentadue, Phys. Lett. {\bf 112B}, 66 (1982).

\bibitem{ADS}
S.M. Aybat, L.J. Dixon, and G. Sterman, Phys. Rev. Lett. {\bf 97},
072001 (2006) [hep-ph/0606254];
Phys. Rev. D {\bf 74}, 074004 (2006) [hep-ph/0607309].

\bibitem{DMS}
L.J. Dixon, L. Magnea, and G. Sterman, JHEP {\bf 08}, 022 (2008) 
[arXiv:0805.3515 [hep-ph]].

\bibitem{BN} 
T. Becher and M. Neubert, Phys. Rev. Lett. {\bf 102}, 162001 (2009) 
[arXiv:0901.0722 [hep-ph]]; 
JHEP {\bf 06}, 081 (2009) [arXiv:0903.1126 [hep-ph]].  

\bibitem{GM}
E. Gardi and L. Magnea, JHEP {\bf 03}, 079 (2009) [arXiv:0901.1091 [hep-ph]];  
Nuovo Cim. C {\bf 32}, 137 (2009) [arXiv:0908.3273 [hep-ph]].

\bibitem{CLS97} 
H. Contopanagos, E. Laenen, and G. Sterman, Nucl. Phys. B {\bf 484}, 303 (1997)
[hep-ph/9604313].

\bibitem{MVV02} 
S. Moch, J.A.M. Vermaseren, and A. Vogt, Nucl. Phys. B {\bf 646}, 181 (2002)
[hep-ph/0209100].

\bibitem{BLS2}
T. Becher, C. Lorentzen, and M.D. Schwartz, Phys. Rev. D {\bf 86}, 054026 
(2012) [arXiv:1206.6115 [hep-ph]]. 

\bibitem{NKBP}
N. Kidonakis and B.D. Pecjak, Eur. Phys. J. C {\bf 72}, 2084 (2012) [arXiv:1108.6063 [hep-ph]].

\end{thebibliography}
\end{document}